\documentclass[aps,pre,twocolumn,10pt,showpacs,showkeys,floatfix]{revtex4-1}
\usepackage{amsfonts}
\usepackage{amssymb}
\usepackage{amsmath}
\usepackage{graphicx}
\usepackage{epsfig}
\usepackage{epstopdf}
\usepackage{color}
\usepackage{bm}
\usepackage{pdfsync}
\usepackage{hyperref} 

\newcommand{\bse}{\begin{subequations}}
\newcommand{\ese}{\end{subequations}}
\newcommand{\be}{\begin{equation}}
\newcommand{\ee}{\end{equation}}
\newcommand{\bea}{\begin{eqnarray}}
\newcommand{\eea}{\end{eqnarray}}
\newcommand{\kb}{k_{_{\mathrm{B}}}}

\newcommand{\chiq}{\chi_{_{\mathrm{q}}}\!}
\newcommand{\chiqp}{\dot{\chi}_{_{\mathrm{q}}}\!}
\newcommand{\chiv}{\chi_{_{\mathrm{v}}}\!}
\newcommand{\chivp}{\dot{\chi}_{_{\mathrm{v}}}\!}

\newcommand{\chiover}{\chi_{_{\mathrm{ov}}}}
\newcommand{\chioverp}{\dot{\chi}_{_{\mathrm{ov}}}}

\newcommand{\qzero}{q_{_{\mathrm{0}}}}

\newcommand{\vzero}{v_{_{\mathrm{0}}}}
\newcommand{\lambdazero}{\lambda_{_{\mathrm{0}}}}
\newcommand{\lambdaf}{\lambda_{_{\mathrm{f}}}}

\newcommand{\lambdap}{\dot{\lambda}}
\newcommand{\tf}{t_{_{\mathrm{f}}}}

\newcommand{\phiv}{\varphi_{_{\!\mathrm{v}}}\!}
\newcommand{\phiq}{\varphi_{_{\!\mathrm{q}}}\!}

\newcommand{\sigmaq}{\sigma_{\mathrm{q}}^{2}}

\newcommand{\phiqover}{\varphi_{_{\!\mathrm{ov}}}}

\newcommand{\qover}{q_{_{\mathrm{ov}}}\!}
\newcommand{\qs}{\widehat{q}(s)}
\newcommand{\fs}{\widehat{f}(s)}
\newcommand{\phis}{\widehat{\varphi}_{_{\mathrm{ov}}}(s)}
\newcommand{\lambdas}{\widehat{\lambda}(s)}
\newcommand{\chis}{\widehat{\chi}_{_{\mathrm{ov}}}(s)}

\allowdisplaybreaks
\begin{document}
%-----------------------------------------------------
\title{Generalized optimal protocols of Brownian motion in a parabolic potential}
\author{Pedro J. Colmenares}
\email{gochocol@gmail.com}
\thanks{Corresponding Author.}
\affiliation{Departamento de Qu\'{\i}mica. Universidad de Los Andes. M\'erida 5101, Venezuela}

\begin{abstract}  
The generalized Langevin equation with an exponential kernel is used to analyze \textcolor{black}{memory effects} on the optimal work done by a Brownian particle in a heat bath and subjected to a harmonic moving potential. The generalized overdamping scenario is also investigated. Several facts emerge in these more precise descriptions using the same initial conditions of the Markovian which lead the particle to do mechanical work against the field. Compared with the results obtained with the latter, the memory fades the discontinuities observed in the highly underdamped regime, which suggests that this trait is a consequence of the Markov approximation as well as the dependence of the different dynamical susceptibilities with the external field. Unlike the overdamped Markovian, work is done by the external field in the analog generalized counterpart. A detailed calculation of the rate of entropy production gives negative values. It is mathematically correct because the dynamics deal with a reduced description of the degrees of freedom of the bath. The theory then requires improving the treatment of them to restore the second law and thus to get the results with the required thermodynamics consistency.  
\end{abstract}
\maketitle

%---------------------------------------------------------
%---------------------------------------------------------
\section{Introduction}
\label{Sec0}
%---------------------------------------------------------
%---------------------------------------------------------

The thermodynamic work involving the interaction of a Brownian particle with a heat bath and an external moving potential has been examined from different perspectives.
From experimental setups where the field was provided by an optical trap displaced at constant speed \cite{ImparatoEtAl,Ciliberto} up to theoretical descriptions where the external protocol minimizes the mechanical work produced by the interaction of the particle with the external field \cite{SchmiedlSeifert2007,AbreuSeifert,PJOscar} have been the mayor researches in the field. In the latter, the analysis has been focused on the Markovian Langevin equation (MLE) in both underdamped and overdamped regimens. However, the description of non-Markovian systems through a more exact description, such as the classical generalized Langevin equation (GLE), has received less attention.  

Although the GLE has been successful in describing non-Markovian systems, there are new results that need to be reconsidered in its formulation. For instance, it can be highlight  the finding by Daldrop {\it et al.} \cite{Daldrop} in the molecular dynamics simulation of a harmonic particle \textcolor{black}{where} the friction is not a constant but depends on the strength of the applied field. \textcolor{black}{It must be mentioned that Vroylandt {\it et al.} \cite{Vroylandt} have also shown that it depends on the position if a non-linear mean force is considered. Then, the general consensus  is that} the field strength must be included in \textcolor{black}{its} interaction with the degrees of freedom of the reservoir \cite{Lisy,DiCairano}. \textcolor{black}{In view of these findings,} there are even sufficient reasons to consider that Kubo's fluctuation dissipation theorem (FDT) associated with the GLE is not so accurate \cite{Lisy,DiCairano}. \textcolor{black}{This deficiency has often been overlooked in the literature, even though it was pointed out decades ago by Balakrishnam \cite{Balakrisnan} and by Cicotti {\it et al} \cite{CicottiRyckaert}. In fact, the work by Olivares-Rivas {\it et al.} \cite{WorPJPhysA} showed that for a constant field the noise correlation depends on its squared strength. \textcolor{black}{Costa {\it et al.} \cite{Costa}} showed that it is even inconsistent in the linear regime of superdiffusive processes.} Additionally, it was also discussed that a dynamics driven by nonequilibrium steady states, the resulting FDT can be written as the equilibrium one and an additive correction \cite{SeifertSpeck2010}. It has been also considered adding correlations between the system and the environment \cite{Kutvonen}.
Despite these basic observations and potentially the derivation of others that are unknown at the moment, there has been some progress using the original GLE in the analysis of the subject matter of this research.

To date, the published works involving the moving harmonic potential incorporate only the memory effect of the stochastic force due to  the reservoir through an exponential kernel as in Ref. \cite{Daldrop}. Thus, Paredes-Altuve {\it et al.} \cite{OscarErnestoPedro} investigated how the individual measurement of the average position or speed affects the optimal protocol and therefore the minimum mechanical work. The no measurement analysis was also included. There, it was shown that the inclusion of memory due to the noise generated by the thermal bath effectively induces the system to perform work against the field when the measurement is done. Furthermore, Di Terlizzi {\it et al.} \cite{Terlizzi} calculated the non-optimal work and the variance of its distribution for optical tweezers displaced at constant speed both in the GLE and in the generalized overdamped version (GOLE).

Under these circumstances, the main objective of this research is to use the classical GLE to analyze memory effects on the determination of  the optimal protocol and minimum work independently of any measurement at all. In particular, to compare the new findings with those obtained from the Markovian case \cite{PJOscar} under the same initial conditions. To do so, the variational procedure of Ref. \cite{OscarErnestoPedro} must be replaced by the more general one implemented in Ref. \cite{PJOscar}. 
It will be of particular interest to find a procedure that, for a given friction coefficient,  a reasonable value of the memory kernel parameter be determined that leads the particle to perform mechanical work against the external field. It will be done for any value of the friction constant assuming that the particle starts from equilibrium and also from an  arbitrary initial position.

To achieve this goal, it will be shown in Sec. \ref{Sec1} the general equations defining the system dynamics and mechanical work. The latter appears to be as a complex functional of the protocol with local and non-local parts \cite{PJOscar}. The solution for the GLE and GOLE is shown in two separate subsections where the work functional is explicitly broken down into its two contributions.  For each one, the associated Euler-Lagrange equation is derived and their sum \cite{FerialdiBassi} is solved to get the optimal driving. It will be shown that for the two regimes, each protocol captures the memory effect of the thermal bath though a well-defined convoluted integral equation. They together, Eqs. (\ref{lambdaGLE}) and (\ref{lambdaGOLE}), are the main results of this proposal. Next, in Sec. \ref{Sec2} is presented the procedure to assign meaningful physical  value of the kernel decay constant and the discussion of the results for the optimal driving and the minimum work. Additionally, the entropy production associated to the dynamics is determined and discussed as ruled by the second law. The article concludes with some general remarks in Sec. \ref{Sec3}.

%---------------------------------------------------------
%---------------------------------------------------------
\section{General equations}
\label{Sec1}
%---------------------------------------------------------
%---------------------------------------------------------
The exact position $q(t)$ of a Brownian particle of mass $M$ submerged in a bath at temperature $T$ and subjected to a moving harmonic potential with stiffness $\kappa^{2}$ and external driving $\lambda(t)$,  that is, $V(q,t)=\kappa^{2}(q-\lambda(t))^{2}/2$, is provided by the GLE \cite{Zwanzig,Banik}:
\be
M\,\ddot{q}(t)=-\!\!\int_{0}^{t}\!\!dy\,\Gamma(t-y)\,q'(y)-\kappa^{2}\left(q(t)\!-\!\lambda(t)\right)+\xi(t),
\ee
where the dot and the apostrophe above a function denotes its time and normal derivative, respectively. 

The kernel $\Gamma(t-y)$ compiles the memory or retardation effect on the movement of the particle due to the collective hydrodynamic response of the bath and $\xi(t)$ is a zero-mean Gaussian colored noise with correlation \textcolor{black}{$\left<\xi(t)\xi(y)\right>=\kb T\,\Gamma(|t-y|)$} satisfying Kubo's FDT \cite{Reichl} where $\kb$ is the Boltzmann constant. 

The analysis presented here will be framed under the guideline of the original Kubo theorem with exponential kernel $\Gamma(t)=\tau\,\exp[-\alpha\,t]$ in order to compare and complement the findings with those of Refs. \cite{OscarErnestoPedro} and  \cite{Terlizzi}, respectively. Constant $\tau=\alpha\,\gamma$ being $1/\alpha$ the kernel decay relaxation time and $\gamma$ the static friction coefficient of the thermal  fluid. 

Scaling time and length by the factors \textcolor{black}{ $\kappa\,M^{-1/2}$} and \textcolor{black}{$\kappa(\kb T)^{-1/2}$}, respectively, the GLE reduces to the dimensionless equation
\bea
\ddot{q}(t)&=&-\tau\!\!\int_{0}^{t}\!\!dy\,f(t-y)\,q'(y)-q(t)+\lambda(t)+\xi(t),
\label{under}
\eea
with $f(t-y)=\exp[-\alpha\,|t-y|]$. Without loss of generality, this is equivalent to assume $\{\kb,\kappa,M\}=1$.

%Making the inertial term equals to zero, the dynamics becomes overdamped with a  characteristic time scale $1/\tau$. Thus, \textcolor{black}{defining the dimensionless $t^{\star}=\tau\,t$ the pair of equations } becomes
%\bea
%\frac{1}{\tau^{2}}\,\ddot{q}(t)&=&-\!\!\int_{0}^{t}\!\!dy\,f(t-y)\,q'(y)-q(t)+\lambda(t)+\xi(t),\label{under}\\
%\int_{0}^{t}\!\!dy\!\!&f&\!\!\!(t-y)\,q'(y)=-q(t)+\lambda(t)+\xi(t),
%\label{over}
%\eea
%espectively, \textcolor{black}{where from now on  the index $\star$ was suppressed}. At this point, we might ask why to choose $\tau$ instead of $\gamma$. The reason lies on the fact that the equations are computationally easier to solve as will be described in Sec. \ref{Sec2}.

The thermodynamic work $W$ has been defined according stochastic energetics in Refs.   \cite{Sekimoto,SeifertReport}.
It is given as a functional of the protocol and its derivative, i.e.
\be
W[\lambda,\lambdap]=\int_{0}^{\tf}dt\,\lambdap(t)\left[\lambda(t)-\left<q[\lambda(t)]\right>\right],
\label{Euler}
\ee
where $\left<q[\lambda(t)]\right>$ is the average over the noise distribution and where it is indicated its explicit functional dependence with the protocol.

The task is to find by means of variational techniques, the optimal $\lambda(t)$ which makes $W[\lambda,\lambdap]$ to reach a minimum taking into account that the functional has a local (``loc'') and non-local (``nl'') contributions. Once they are identified, the first variation is
\be
\delta W_{r}[\lambda,\lambdap]\!=\!\!\int_{0}^{\tf}\!\!\!\!\!dt\!\left(\!\frac{\partial  L_{\mathrm{r}}[\lambda,\dot{\lambda},t]}{\partial\lambda}-\!\frac{d}{dt}\frac{\partial  L_{\mathrm{r}}[\lambda,\dot{\lambda},t]}{\partial\dot{\lambda}}\!\right)\!\delta\lambda(t),
\label{Wlnl}
\ee
where $\mathrm{r}=\{\mathrm{loc},\mathrm{nl}\}$ and $ L_{\mathrm{r}}[\lambda,\lambdap,t]$ is a definite function to be determined. The first variation $\delta W$ is then given as the sum of the two associated Euler-Lagrange equations \cite{FerialdiBassi,PJOscar} which, after setting it equal to zero will allow to determine the optimal protocol $\lambda(t)$.

Once the protocol and the average position are derived then $W$ can be integrated between $\{0,\tf\}$ where $\tf$ is the final time of the protocol application. It renders,
\be
W=\frac{1}{2}\lambdaf^{2}-\lambdaf\left<q(\tf)\right>+\int_{0}^{\tf}dy\,\lambda(y)\left<q'[\lambda(y)]\right>),
\label{Work}
\ee
where $\lambdaf=\lambda(\tf)$ is a pre-fixed values fixed by the external agent. We have supposed as in Ref. \cite{PJOscar} that $\lambda(0)=0$. The first two terms are just the adiabatic work by changing the protocol to $\lambdaf$ and instantly returning it to the initial value.

The determination of $q(t)$ for the GLE and for the generalized overdamped and also their associated optimal protocol which makes the work minimal, will be shown in the next two subsections.

%---------------------------------------------------------
%---------------------------------------------------------
\subsection{GLE dynamics}
%---------------------------------------------------------
%---------------------------------------------------------
The Laplace transform of Eq. (\ref{under}) gives
\bea
\widehat{q}(s)&=&\qzero\widehat{\chiq}(s)+\vzero\,\widehat{\chiv}(s)+\widehat{\chiv}(s)\,\widehat{\lambda}(s)+\widehat{\phiq}(s),\label{qs}\\
\widehat{\chiv}(s)&=&\textcolor{black}{\frac{1}{s\left(s+\tau\widehat{f}(s)\right)+1},}\label{chivunder}\\
\widehat{\chiq}(s)&=&\textcolor{black}{\left(s+\tau\widehat{f}(s)\right)\widehat{\chiv}(s)},\label{chiqunder}\\
\widehat{\phiq}(s)&=&\widehat{\chiv}(s)\,\widehat{\xi}(s),
\eea
where $\vzero$ and $\qzero$ are the initial velocity and position, respectively. Functions $\widehat{\chiv}(s)$ and $\widehat{\chiq}(s)$ are the so-called susceptibilities.

Inverting Eq. (\ref{qs}) and averaging with the Maxwell velocity distribution gives the solution of the GLE for any value of $\tau$,
\bea
q(t)&=&\left<q[\lambda(t)]\right>+\phiq(t),\label{qunder}\\
\left<q[\lambda(t)]\right>)&=&\qzero\,\chiq(t)+\textcolor{black}{\int_{0}^{t}dy\,\chiv(t-y)\,\lambda(y)},\label{qbar}\\
\phiq(t)&=&\textcolor{black}{\int_{0}^{t}dy\,\chiv(t-y)\,\xi(y)}\label{phiunder}. 
\eea

\textcolor{black}{The optimal protocol is obtained by minimizing Eq. (\ref{Euler}) with respect to $\lambda(t)$. This is accomplished by substituting it into the work equation, Eq. (\ref{qunder}), to obtain a functional both local, where its value at a given moment depends only on that point, and non-local through its integration in the considered time interval. Each contribution has its associated Euler-Lagrange equation (ELE) that according to Eq. (\ref{Wlnl}) gives the first variation of $W[\lambda(t)]$ in terms of its functional derivative. Setting it to zero allows the optimal protocol to be determined. It was already addressed in the appendix of the previous work \cite{PJOscar} for an equation similar to Eq. (\ref{qunder}) that obeys a dynamic given by the classical Langevin equation}. Thus, fitting \textcolor{black}{the mentioned procedure to the GLE}, the \textcolor{black}{first functional derivative} associated to Eq. (\ref{Wlnl})  \textcolor{black}{given by,} 

\bea
\int_{0}^{t}dy\,\lambda(y)\,A_{_{4}}(t,y)&=&-A_{_{1}}(t)+A_{_{2}}(t)\nonumber\\
&-&\int_{0}^{\tf}dy\, \lambda(y)\,A_{_{3}}(t,y),\\
A_{_{1}}(t)&=&\qzero\,\chiqp(t),\\
A_{_{2}}(t)&=&\textcolor{black}{\lambdaf\,\chiv(\tf-t)},\\
A_{_{3}}(t,y)&=&\textcolor{black}{\chiv'(y-t)},\\
A_{_{4}}(t,y)&=&\textcolor{black}{\left[\chivp(t-y)-\chiv'(y-t)\right]}\!.
\eea

Taking the Laplace transform of the convolution and inverting gives the desired expression for $\lambda(t)$, that is,
\bea
\lambda(t)&=&\phi_{_{1}}(t)-\int_{0}^{\tf}dy\,\phi_{_{2}}(t,y)\,\lambda(y),\label{lambdaGLE}\\
\phi_{_{1}}(t)&=&\mathcal{L}^{-1}\left[\frac{\widehat{A}_{2}(s)-\widehat{A}_{1}(s)}{\widehat{A}_{5}(s,t)}\right],\\
\phi_{_{2}}(t,y)&=&\mathcal{L}^{-1}\left[\frac{\widehat{A}_{3}(s,y)}{\widehat{A}_{5}(s)}\right],\\
\widehat{A}_{5}(s)&=&s\left(\widehat{\chiv}(s)-\mathcal{L}\left\{\chiv(-t)\right\}\right).
\eea

Equation \ref{lambdaGLE} incorporates the whole history of the dynamics when the field is turned on.
Functions $\phi_{_{i}}(t)$ are well-defined but unfortunately are long expressions to show them explicitly. They compile together the effect of the memory kernel on the protocol. Fortunately, it appears to be defined as a Fredholm integral equation of the second kind for which the Newman method \cite{Arfken} is suitable to solve it using $\phi_{_{1}}(t)$ as the seed of the iteration process. The convergence is reached after a convenient number of steps as it will be shown in Sec. \ref{Sec2}. This is the first mayor result.

Once Eqs. (\ref{lambdaGLE} is numerically solved and substituted back into Eq. (\ref{qbar}), then the mechanical work can be obtained from Eq. (\ref{Work}). 
%---------------------------------------------------------
%---------------------------------------------------------
\subsection{GOLE dynamics}
%---------------------------------------------------------
%---------------------------------------------------------
\textcolor{black}{The generalized overdamped equation is obtained by setting $\ddot{q}(t)=0$ in Eq. (\ref{under}). Its solution} can be found in a similar way as for the GLE by transforming it to the Laplace space. It gives $\qs=\textcolor{black}{\chis(\qzero\,\tau\,\fs+\lambdas+\phis)}$ with $\chis=\textcolor{black}{1/(\tau\,s\,\fs+1})$ being the corresponding susceptibility. From the latter $\textcolor{black}{\tau\,\fs\chis}=\left(1-\chis\right)/s$ which after substituting it into $\qs$ and inverting finally renders
\bea
\qover(t)&=&\left<\qover[\lambda(t)]\right> +\varphi(t),\label{qsover}\\
\left<\qover[\lambda(t)]\right>&=&\qzero\left(1-\chi(t)\right)\nonumber\\
&+&\int_{0}^{t}dy\,\chiover(t-y)\,\lambda(y),\label{qoverp}\\
\varphi(t)&=&\int_{0}^{t}dy\,\chiover(t-y)\,\xi(y),\label{phiover}\\
\chiover(t)&=&\textcolor{black}{\mathcal{L}^{-1}\left[\frac{1}{\tau\,s\,\widehat{f}(s)+1}\right],\label{chiover}}\\
\chi(t)&=&\int_{0}^{t}dy\,\chiover(y).
\eea
This solution, however,  has an associated stationary PDF inconsistent with the Boltzmann-Gibbs distribution as was proved by Nascimiento and Morgado \cite{Nascimiento}. The reason lies on the physical consequences of the the memory kernel on the dynamics. Physically, it gives an account of the particle movement through the flow lines of the bath fluid. However, for \textcolor{black}{overdamped} systems the inertia is irrelevant since the particle velocity relaxation is very fast reaching the same of the  reservoir. The interaction of particle with the bath is almost instantaneous which is not considered in the GLE by merely setting the inertial term equals to zero. In order to correct this physical deficiency, these authors proposed to add an extra white noise $\eta(t)$  with intensity $\gamma_{_{0}}$ and correlation $\left<\eta(t)\eta(y)\right>=2\gamma_{_{0}}T\delta(t-y)$ to the non-inertial GLE. In this circumstance, GOLE's correct noise must be
\be
\phiqover(y)=\int_{0}^{t}dy\,\chiover(t-y)\left(\xi(t)+\eta(t)\right),\label{corr}
\ee
so that according to the FDT and that Eq. (\ref{qsover}) reaches the Boltzmann-Gibbs distribution at equilibrium, the extra term   $\gamma_{_{0}}\,\delta(t)$ must be added to $f(t)$ \cite{Nascimiento}. 
 These arguments supported by a detailed mathematical analysis  were subscribed by Di Terlizzi {\it et al.}\cite{Terlizzi} in their GOLE assuming the optical trap is moved at constant rate. They used $2\,\gamma_{_{0}}\,\delta(t)$ instead to give account of the experiment by Berner {\it et al.} in viscoelastic fluids \cite{Berner}. 
 
Then, assuming the particle is rigid,  replacing Eq. (\ref{qoverp}) into Eq. (\ref{Euler}) and recalling that $\chiover(t)$ is determined from a kernel given by the sum of $\gamma_{_{0}}\,\delta(t)$ plus the exponential, the corresponding generalized overdamped functions appearing in Eq. (\ref{Wlnl}) are given by:
\bea
 L_{\mathrm{loc}}(\lambda,\lambdap,t)&=&\lambdap(t)\lambda(t)-\qzero\!\left(1-\chi(t)\right)\lambdap(t),\\
  L_{\mathrm{nl}}(\lambda,\lambdap,t)&=&-\lambdap(t)\!\int_{0}^{t}dy\,\chiover\!(t-y)\,\lambda(y).
 \eea

Thus, the local part of $\delta W$ is simply
\bea
\delta W_{\mathrm{loc}}&=&\int_{0}^{\tf}dt\,\lambdap(t)\,\delta\lambda(t)\nonumber\\
&+&\int_{0}^{\tf}\,dt\left(\lambda(t)-\qzero\left(1-\chi(t)\right)\right)\delta\lambdap(t),
\eea
where the functional dependence was omitted for shortness.

Similarly, the non-local contribution is \textcolor{black}{obtained using the general guidelines of the procedure applied to the underdamped. It reads}
\bea
\delta W_{\mathrm{\!nl}}&=&-\int_{0}^{\tf}dt\,\delta\lambda(t)\int_{t}^{\tf}dy\,\chiover(y-t)\,\lambda'(y)\nonumber\\
&-&\int_{0}^{\tf}dt\,\delta\lambdap(t)\int_{0}^{t}dy\,\chiover(t-y)\,\lambda(y).
\eea

Setting to zero the sum of the associated  Euler-Lagrange equations of these two contributions and splitting the integral involving the limits $\{t,\tf\}$ gives the final result
\bea
\int_{0}^{t}dy\,A_{_{1}}(t,y)\!&-&\!\int_{0}^{\tf}\!\!dy\,A_{_{2}}(t,y)\lambda'(y)\!+\!A_{_{3}}(t)=0,\label{lover}\\
A_{_{1}}(t,y)&=&\lambda'(y)\,\chiover(y\!-\!t)\!+\!\lambda(y)\,\chioverp(t\!-\!y),\\
A_{_{2}}(t,y)&=&\chiover(y-t),\\
A_{_{3}}(t)&=&\qzero\,\dot{\chi}(t)+\chiover(0)\,\lambda(t).
\eea
To proceed further, the Laplace transform of Eq. (\ref{lover}) renders
\bea
\widehat{\lambda}(s)&=&\widehat{\Psi}_{_{1}}(s)+\int_{0}^{\tf}dy\,\widehat{\Psi}_{_{2}}(s,y)\,\lambda'(y),\label{psis}\\
\widehat{\Psi}_{_{1}}(s)&=&\frac{\lambdazero\,\widehat{\chi}_{_{\mathrm{ov}}}^{\,\star}(s)-\qzero\left(s\,\chi(s)-\chi(0)\right)}{\widehat{A}_{_{4}}(s)},\\
\widehat{\Psi}_{_{2}}(s,y)&=&\frac{\widehat{A}_{2}(s,y)}{\widehat{A}_{4}(s)},\\
\widehat{A}_{_{4}}(s)&=& s\left(\widehat{\chi}_{_{\mathrm{ov}}}^{\,\star}(s)+\widehat{\chi}_{_{\mathrm{ov}}}(s)\right),\\
\widehat{\chi}_{_{\mathrm{ov}}}^{\,\star}(s)&=&\mathcal{L}\{\chiover(-t)\}.
\eea

Inverting Eq (\ref{psis}) finally gives 
\be
\lambda(t)=\Psi_{_{\!\!1}}(t)+\int_{0}^{\tf}dy\,\Psi_{_{\!\!2}}(t,y)\,\lambda'(y),\label{lambdaGOLE}
\ee
with the two functions $\Psi_{_{\!\!1}}(t)$ and $\Psi_{_{\!\!2}}(t,y)$ given by \cite{Math}:
\begin{widetext}

\bea
\Psi_{_{\!\!1}}(t)&=&\frac{1}{2}\left[\lambdazero-\qzero+\left(\qzero+\lambdazero\right)\left(\frac{\alpha}{\omega_{_{1}}}t+\frac{\left(1+\omega_{_{1}}^{2}\right)}{\omega_{_{1}}^{2}\,\sqrt{\alpha\,\gamma_{_{0}}}}\,\sinh\left(\sqrt{\frac{\alpha}{\gamma_{_{0}}}}\,\omega_{_{1}}\,t\right)\right)\right],\\
\Psi_{_{\!\!2}}(t,y)&=&\frac{\exp \left[-\frac{(1+\omega_{_{1}}^{2})}{2\,\gamma_{_{0}}}y\right]}{2\,\sqrt{\alpha\,\gamma_{_{0}}}\,\omega_{_{1}}^{2}\,\omega_{_{2}}}
\Bigg\{\cosh\left(\frac{\omega_{_{2}}}{2\,\gamma_{_{0}}}y\right)\left[\omega_{_{1}}\,\omega_{_{2}}\,\sqrt{\alpha\,\gamma_{_{0}}}\left(\omega_{_{1}}^{2}+\alpha\,t\right)+\left(1+\omega_{_{1}}^{2}\right)^{2}\sinh\left(\sqrt{\frac{\alpha}{\gamma_{_{0}}}}\,\omega_{_{1}}\,t\right)
  \right]\nonumber\\
  &+&\left.\sinh\left(\frac{\omega_{_{2}}}{2\,\gamma_{_{0}}}y\right)\left[\sqrt{\alpha\,\gamma_{_{0}}}\,\,\omega_{_{1}}\,\left(1+\omega_{_{1}}^{2}+\alpha^{2}\,\gamma_{_{0}}\left(\gamma_{_{0}}+t\right)\right)-2\,\sqrt{\alpha\,\gamma_{_{0}}}\,\,\omega_{_{1}}\left(1+\omega_{_{1}}^{2}\right)\cosh\left(\sqrt{\frac{\alpha}{\gamma_{_{0}}}}\,\omega_{_{1}}\,t\right)\right.\right.\nonumber\\
 &-&\left(1+\omega_{_{1}}^{2}\right)^{2}\sinh\left(\sqrt{\frac{\alpha}{\gamma_{_{0}}}}\,\omega_{_{1}}\,t^{2}\right) \bigg]\Bigg\},
\eea
\end{widetext}
where $\omega_{_{1}}=\sqrt{1+\alpha\,\gamma_{_{0}}}$ and $\omega_{_{2}}=\sqrt{4+\alpha^{2}\,\gamma_{_{0}}^{2}}$. The integral equation (\ref{psis}) converges after a moderate number of iterations if $\Psi_{_{\!\!1}}'(y)$ is used as the initial guess for $\lambda'(y)$. 
Functions $\left<\qover[\lambda(t)]\right>$, $\left<\dot{q}_{_{\mathrm{ov}}}\![\lambda(t)]\right> $ and $W$ are long expressions to display so they are  numerically determined.

In the next section, it will be presented an auto-consistent procedure to calculate the optimal protocol and minimum work for GLE and GOLE. \textcolor{black}{They will be compared with those obtained for the same system using the MLE  \cite{PJOscar} as a standard or point of reference.} Likewise, it will discuss the results in a thermodynamic basis. 

%---------------------------------------------------------
%---------------------------------------------------------
\section{Discussion of results}
\label{Sec2}
%---------------------------------------------------------
%---------------------------------------------------------
The chosen initial conditions are the same as in Ref. \cite{PJOscar} where the Markovian equations lead to a negative mechanical work. Namely, $\{\qzero,\lambdaf\}=\{1,0\}$. The final time of the protocol application $\tf$ is 12 and the friction coefficient for the GLE ranges from the very underdamped $\gamma=1$ up to the overdamped  described by  GOLE. The decay constant $\alpha$ in the GLE is assumed to decrease with the friction to incorporate intuitively the retardation in the particle translation as $\gamma$ gets higher. It was chosen to get $W$ of the same order of magnitude for the considered initial set and all $\gamma$ values. It should be mentioned that the GLE algorithm is unstable for high $\tau$ due to high oscillations in the functions $\phi_{1}(t)$ and $\phi_{2}(t)$ and because of it, it is replaced for that of the GOLE.  Additionally, the selected $\alpha$ for a given $\gamma$ is critical otherwise the oscillations mentioned before show unbound values. 

The plots were calculated for increasing values of $\tau$ and segmented in terms of $\gamma$ and $\alpha$. The $\alpha$ for a given $\gamma$ was arbitrarily chosen in such a way that it generate mechanical work done by the particle in the first three cases, $\gamma=\{1,3,6\}$  and done by the field in GOLE.  Besides the increasing $\tau$ condition, an extra restriction was used to tune up the selected $\alpha$. It is related to the total work involved over the entire time range such that for the assigned decay constant, this quantity stays around a fixed meaningful value. It was fixed to be around \textcolor{black}{$ -0.94$} for $\gamma=\{1,3,6\}$ while for GOLE was around 15. In this way, the color tags used in the plots are for $\gamma$ of 1 (black), 3 (red), 6 (blue) and GOLE (green) with their respective associated $\alpha$ values \textcolor{black}{$\{0.16,0.13,0.09\}$} for the first three cases. \textcolor{black}{For GOLE a high $\gamma$ and a low $\alpha$ is required, thus we set $\alpha=0.001$, $\gamma=1/\alpha$ and $\gamma_{_{0}}=5$}. 
Using the above procedure to select $\alpha$, the numerical determination of $\lambda(t)$ and the desired $W$  is very efficient in the GLE.

It is important to mention that GOLE's calculations are very sensitive to \textcolor{black}{ the value of $\gamma_{_{0}}$}. Convergence is reached for a given \textcolor{black}{value in such a way that} the memory effect avoids taking over the non-inertial dynamics. For other values besides this critical \textcolor{black}{value}, the solution is out of the prescribed bounds with no physical sense at all.

Since no physical arguments have been used to assign the correct value of the decay constant, computer simulation seems to be the only way to get the proper information about its magnitude. Various methods could be used to accomplish this purpose. The first could be the technique developed by Ceriotti {\it et al.} \cite{Ceriotti} to use the GLE to perform an isothermal molecular dynamics simulation and the second,  that from Stella {\it et al.} \cite{Stella} where the colored exponential noise is consistently generated.

\begin{figure}[h]
\centering
\includegraphics[width=1.4\linewidth]{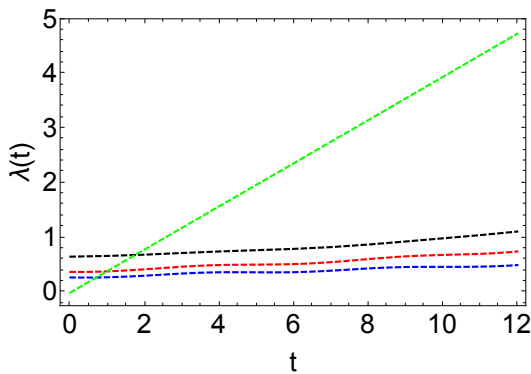}
\vspace{-9.5cm}
\caption{\textcolor{black}{Time- dependent optimal protocol in terms of their defining parameters $\{\gamma,\alpha\}$ of $\{1,0.16\}$ (black), $\{3,0.13\}$ (red) and $\{6,0.09\}$ (blue). The green curve is for GOLE with $\alpha=0.001$ and $\gamma=1/\alpha$ to fulfill the non-inertial character of the description. The set $\{\qzero,\lambdaf\}=\{1,0\}$. See text for details.}}
\label{Fig1}
\end{figure}

Based on this reasoning, the optimal driving are depicted in Fig. \ref{Fig1}. 

\vspace{-1.5cm}

\textcolor{black}{For the underdamped, they are non-linear functions of time unlike the straight lines predicted in the Markovian case \cite{PJOscar}.  GOLE shows an opposite behavior in comparison with the overdamped MLE where $\lambda(t)$ decreases with time. Together, they have} a non-vanishing value at $t=0$ not shown because of the scale. 

\begin{figure}[ht]
\centering
\includegraphics[width=1.4\linewidth]{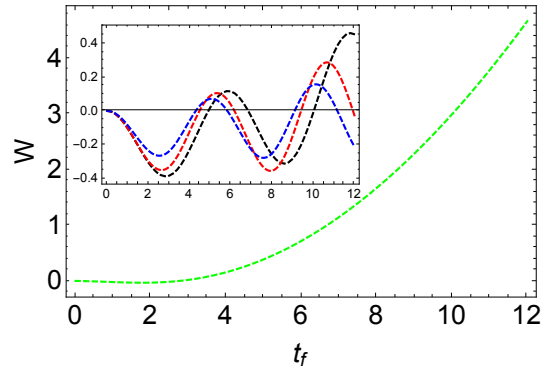}
\vspace{-9.9cm}
\caption{Minimum work for the protocols of Fig. \ref{Fig1}. }
\label{Fig2}
\end{figure}

The minimum work is shown in Fig. \ref{Fig2} for the optimal protocols of Fig. \ref{Fig1}.

%\vspace{-1.3cm}

The differences between the results shown in this graph compared to those obtained with the MLE counterpart become now more than evident. \textcolor{black}{Particularly, there is no observation of discontinuities in the very high underdamped as seen in Fig. 1 of  Ref. \cite{PJOscar} with the $\gamma=1$ curve (black).  The similarity in behavior of the GLE underdamped curves is due to the requirement that $W$ has a fixed value around $-0.94$ throughout the time interval.} Furthermore, excluding the overdamped, GLE does not predict a smooth and negative $W$ as in MLE, but has an almost periodic-like behavior with positive values in some ranges of the protocol application time. Extractable work of  these characteristics were also found by Bylicka {\it et al.} \cite{Bylicka} in open quantum systems with memory effects. Furthermore, GOLE predicts \textcolor{black}{a behavior contrary} to the Markovian.

One way to analyze it is through the Fokker-Planck formalism. It requires knowing the reduced conditional probability distribution function  (PDF), $p(q,t|\qzero)$. The method to obtain it is fully described in Secs. 2 y 3.1 of Ref. \cite{WorPJPhysA}. Making it suitable to the problems at hand, it is a Gaussian given by

\bea
p(q,t|\qzero)&=&\frac{1}{\sqrt{2\,\pi\,\sigma^{2}(t)}}\exp\left[-\frac{\left(q-H(t)\right)^{2}}{2\,\sigma^{2}(t)}\right]\label{pfpe},
\eea
where the standard deviation $\sigmaq(t)$ and function $H(t)$ are defined  respectively as:
\bea
& &\hspace{1cm}\mathrm{GLE}\nonumber\\ 
\sigma^{2}(t)&=&2\int_{0}^{t}\!\!\!\!dy\int_{0}^{y}\!\!\!\!dz\,\left<\phiv(y)\phiv(z)\right>+T\chi_{_{\mathrm{v}}}^{2}(t),\\
H(t)&=&\left<q[\lambda(t)]\right>),\\
\left<\phiv(t)\phiv(s)\right>&=&2\,\gamma\,\alpha\, T\,\int_{0}^{t}dy\,\chivp(t-y)\nonumber\\
&\times&\int_{0}^{s}dz\,\chivp(s-z)\,f( |y-z|),\nonumber\\
& &\hspace{1cm}\mathrm{GOLE}\\ 
\sigma^{2}(t)&=&2\int_{0}^{t}dy\int_{0}^{y}dz\,\left<\phiqover(y)\phiqover(z)\right>,\\
H(t)&=&\left<\qover[\lambda(t)]\right>
\eea
The correlation $\left<\phiqover(t)\phiqover(s)\right>$ can be determined  assuming the noises are uncorrelated with $\left<\eta(t)\eta(s)\right>=\gamma_{_{0}}\,\delta(t-s)$ and $\left<\xi(t)\xi(s)\right>=f(|t-s))$.

\begin{figure}[ht]
%\centering
\includegraphics[height=17cm,width=15cm,angle=0]{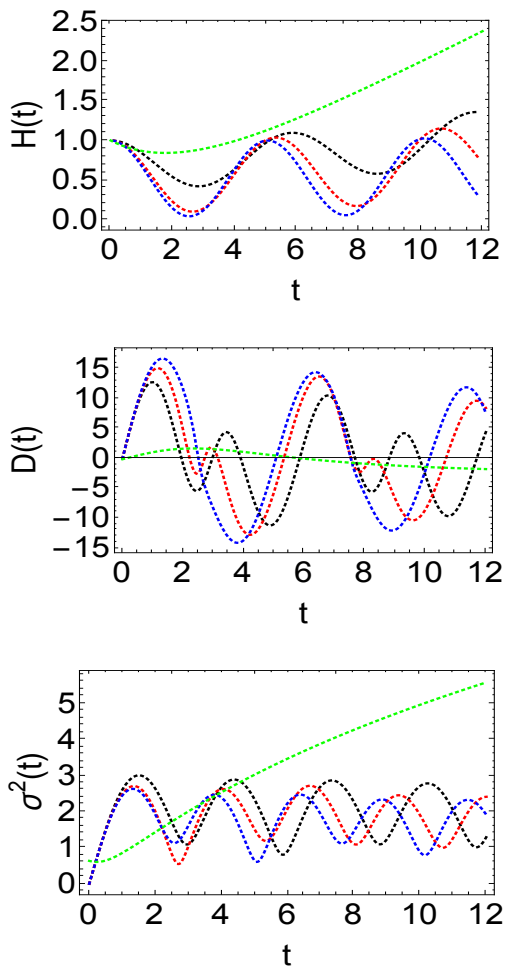}
\vspace{-2.4cm}
\caption{Function $H(t)=\left<q(t)\right>$, effective diffusion constant $D(t)$ and standard deviation $\sigma^{2}(t)$ for GLE and GOLE. Color scheme as of Fig. \ref{Fig1}.}
\label{Fig3}
\end{figure}

The Fokker-Planck equation whose solution is the above Gaussian is determined by applying the procedure originally designed by Adelman and Garrison \cite{Adelman1} and detailed shown in Ref. \cite{PJPRE}. It renders,
\bea
  \frac{\partial p(q,t)}{\partial t}&=&-\frac{\partial J(t,q)}{\partial q},\\
  J(q,t)&=&\Omega(t)\,q\,p(q,t)-\frac{1}{2}D(t)\frac{\partial p(q,t)}{\partial q},\label{J}\\
  \Omega(t)&=&\frac{\dot{H}(t)}{H(t)},\\
  D(t)&=&\dot{\sigma}^{2}(t)-2\,\sigma^2(t)\,\Omega(t),
  \eea
where $D(t)$ is the effective diffusion constant. It is shown in Fig. \ref{Fig3} \textcolor{black}{for $T=10$} along with the average position and standard deviation.
%\vspace{-1cm}

Regarding the average position $H(t)$, we see that \textcolor{black}{ all curves tend to diverge from the potential center being more evident in GOLE} due to the instantaneous hits received from the particles of the thermal bath. This is also manifested in the protocols shown in Fig. \ref{Fig1} through the final jumps. 

\textcolor{black}{The diffusion constant appears to be a smooth function for all the considered friction constants unlike the strong Markovian underdamped which shows discontinuities in $D(t)$ because of the ratio  $\Omega(t)$. This was extensively discussed in a previous paper \cite{PJPRE2} for the Markovian harmonic oscillator. The effect of average velocity and position on the above relationship makes the GLE's $D(t)$ a continuous function of time. Then we can conclude that the MLE applied to the highly underdamped case cannot cope with the cusps in $D(t)$ because the averages describe a physical situation without the translational memory. This means that the mechanical work has no clear physical meaning because it also has unexpected cusps \cite{PJOscar}.}

The standard deviation $\sigmaq(t)$ results of the combined effects of the noise correlation function, the effective diffusion constant and the ratio $\Omega(t)$.  Its non-monotonicity is primarily due to the effects of the oscillations shown by the susceptibilities and the functions $\varphi_{1}(t)$ and $\varphi_{2}(t,y)$. They are canceled out in the overdamped giving a linear function because of the preeminence of the intensity of white noise over the memory kernel $f(t)$. 
What is important to notice is that GOLE being a more accurate theory, \textcolor{black}{it gives a totally different $\sigma^{2}(t)$ compared with the saturated  curve of the MLE description \cite{PJPRE2}. }

%---------------------------------------------------------
%---------------------------------------------------------
\subsection{Entropy production rate}
\label{Sec2a}
%---------------------------------------------------------
%---------------------------------------------------------

To give an account of the thermodynamic cost invested in doing work by or against the field, entropy $S=-\int dq\,p(q,t|\qzero)\,\ln[p(q,t|\qzero)]$ is the proper property to be called upon. This property was detailed analyzed by Seifert along the trajectory of the Brownian particle  \cite{Seifert} and subsequently by van den Broeck and Esposito in terms of the processes responsible for the breaking of the detailed balance once the protocol is initiated \cite{BroeckEsposito}.

\begin{figure}[h]
 \includegraphics[height=16cm,width=14cm,angle=0]{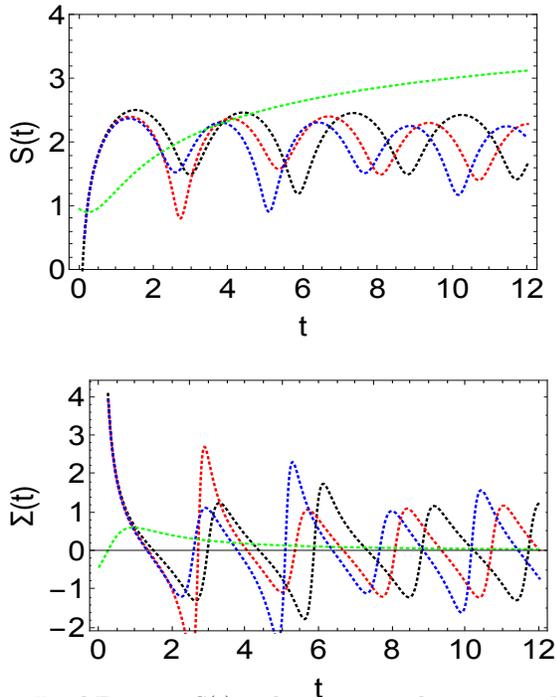}
\vspace{-6cm}
\caption{Total Entropy $S(t)$ and entropy production rate $\Sigma(t)$ for GLE and GOLE probability distributions. Color scheme as of Fig. \ref{Fig1}.}
\label{Fig4}
\end{figure}

The standard deviation is crucial on the behavior of the total entropy $S(t)$ and its derivative, the entropy rate $\Sigma(t)$. They can be easily determined from Eq. (\ref{pfpe}) and plotted in Fig. \ref{Fig4}.

 %\vspace{-1.cm}

It is observed a negativity temporarily character of $\Sigma(t)$. We must emphasize, however, they are legitimate outcomes own of the non-Markovian dynamics despite of the provisions of the Second Law. In fact, because the bath degrees of freedom (BDF) in the theory  are not explicitly considered but have been \textcolor{black}{described by the static friction coefficient},  their real contribution to the entropy balance are excluded so their explicit incorporation to the total will provide the required positivity.  This kind of ``thermodynamical inconsistent'' behavior is inherent to non-Markovian systems. It was proved by Marcantoni {\it et al.} \cite{Marcantoni} for open quantum systems described by non-Markovian maps and by Strasberg and Esposito \cite{StrasbergEsposito} for a dynamics based on an effective master equation where the non-Markovian character is incorporated through an transition matrix depending on the initial state having negative entries and fulfilling the Chapmann-Kolmogorov equation. Moreover, Bonan\c{c}a {\it et al.} \cite{Bonanca} have shown for the electrical conduction in metals that temporal negative entropy production also spring out as a direct consequence of Ohm's law. In the language of the GLE, the Ohmic driving of the dynamics is the non-Markovian dissipation term \cite{SchrammJungGrabert}. 

Therefore, it is plausible to assume that the control of the thermodynamic outcome by the standard deviation shown in Fig. \ref{Fig3} along with the time-dependent diffusion term, are responsible for the system showing a negative rate of entropy production. This property will be further analyzed next. 

In a system in equilibrium, the correlation of the fluctuations for small external disturbances is given in terms of the response function. This is not the case in stationary systems out of equilibrium because the detailed balance relation is broken and therefore there is a continuous degradation of energy to the thermal reservoir. This behavior occurs along the trajectory which in turn reduces to the conventional FDT if the velocity fluctuations are given in terms of the local mean velocity \cite{SpeckSeifert3}. 

In the detailed balance restoration, van den Broeck and Sposito \cite{BroeckEsposito} claim that entropy production involves the interplay of two different mechanisms. One is the adiabatic (ad) recovering of the temporal symmetry due to any given constraint and the second one, the persistent non-adiabatic (na) influence of the field thought the driving. 

 The two mentioned contributions to the entropy rate are given by Eqs. (27) and (28)  of Ref. \cite{BroeckEsposito}, which applied to Eq. (\ref{J}) gives,
  \bea
  \Sigma_{\mathrm{na}}(t)&=&\frac{2}{D(t)}\int_{-\infty}^{\infty}dq\,p(q,t|\qzero)\left(\frac{J(q,t)}{p(q,t|\qzero)}\right)^{\!2},\\
  &=&\frac{D(t)}{2\,\sigma^{2}(t)}\!+\!2\,\Omega(t)\!\left(\!1\!+\!\Omega(t)\frac{H^{2}(t)\!+\!\sigma^{2}(t)}{D(t)}\right)\!\!,\\
 \Sigma_{\mathrm{ad}}(t)&=&0.
  \eea
  
There is not an adiabatic contribution since the term  $J^{\mathrm{st}}(q,t)/p^{\mathrm{st}}(q,t)$ vanishes because of the stationary (st) PDF
  \be
  p^{\mathrm{st}}(q,t)=\sqrt{-\frac{\Omega(t)}{\pi\,D(t)}}\exp\left[\frac{\Omega(t)}{D(t)}q^{2}\right].
  \ee
  The negativity of $\Omega(t)/D(t)$ is guaranteed so the system reaches the steady state. It is important to recall that $\Sigma_{\mathrm{na}}(t)$ agrees with that by Seifert \cite{Seifert} in which the rate is calculated along the trajectory.
  
 Hence, the protocol  is the only source of entropy production to restore the symmetry breaking since there are no constrictions of any kind on the dynamics.
 
\begin{figure}[h]
\includegraphics[width=1.4\linewidth]{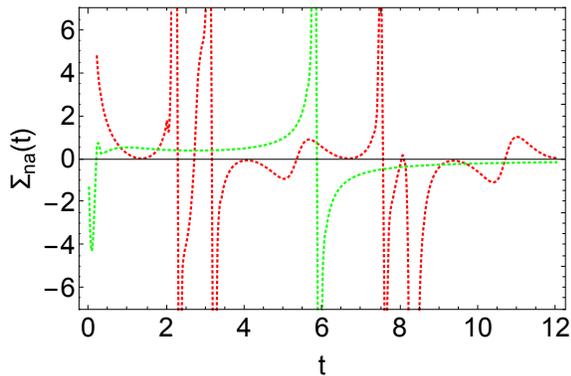}
\vspace{-9.5cm}
\caption{\textcolor{black}{Non-adiabatic entropy production rate $\Sigma_{\mathrm{na}}(t)$ for $\gamma=3$ and GOLE. }}
\label{Fig5}
\end{figure}

\textcolor{black}{The entropy rate $\Sigma_{\mathrm{na}} (t)$ is a complicated function of time showing sudden changes at given times for all values of the friction constant. Furthermore, $\int_{0}^{\tf}dt\,\Sigma_{\mathrm{na}}(t)$ is positive. These two facts are due to the non-rigorous entropy balance from the effective management of the BDFs.
For the matter of presentation, it is only shown in Fig. \ref{Fig5} the results for $\gamma=3$ and GOLE. }

%\vspace{-1.5cm}

It can be seen, that indeed, the temporally negative entropy production rates in discrete non-Markovian systems also appear in the non-Markovian Langevin equation in both the inertial and underdamped descriptions. 

In order to test this result then it is important to determine the entropy production, $\Sigma_{\mathrm{ex}}(t)$,  due to the heat dissipated to the bath. Using any of the approaches  \cite{Seifert,BroeckEsposito} it is found that
\bea
\Sigma_{\mathrm{ex}}(t)&=&-2\frac{\Omega(t)}{D(t)}\int_{-\infty}^{\infty}dq\,q\,J(q,t),\\
&=&-\frac{\Omega(t)}{D(t)}\left(D(t)\!+\!2\,\Omega(t)\left(H^{2}(t)\!+\!\sigma(t)\right)\right)\!,
\eea
which added to $\Sigma_{\mathrm{na}}(t)$ gives $\Sigma(t)$ as it should be.  

These results do not give any clues about the effects that the actual production of entropy has on the ability of the system to do work against the external field. The negative entropy rate is a consequence of the ineffective management of the bath degrees of freedom requiring a reformulation of the FDT as was mentioned in the introduction. In particular, the observation by Kutvonen  {\it et al.} \cite{Kutvonen} that Markov limit of non-Markovian systems leads to additional entropy components. \textcolor{black}{The absence of it in the MLE can provide a secondary route to thermodynamically understand the cusp appearing in the mechanical work mentioned above. Recently, Hsiang and Hu \cite{Hsiang} derived the FDT for systems subjected to non-equilibrium dynamics driven by an external protocol.}

Despite these arguments, the results provided by GLE and GOLE are mathematically correct. It corroborates the work of Van den Broeck and Esposito, where a dissipative-driven dynamics of coarse-grained non-Markov systems, the rate of entropy production is naturally negative \cite{StrasbergEsposito}.

However, since the classical GLE is only an approximation of a real system based on an FDT framed in linear response theory, then it is plausible to assume that a "more consistent" formulation of it will provide results to be in accordance with the prescriptions of the second law. 

%---------------------------------------------------------
%---------------------------------------------------------
\section{Concluding remarks}
\label{Sec3}
%---------------------------------------------------------
%---------------------------------------------------------

Summarizing, the auto-consistent inclusion of memory effects definitely changes the expected value of the work done by the particle in comparison with the Markovian counterpart \cite{PJOscar}. The initial position other than the equilibrium values is responsible for the particle to do work against the external field except in the generalized overdamped. The calculations for other initial conditions will show as in \cite{PJOscar} that work is done by the field. The generalized overdamped protocol is, as in the Markovian, a general result no matter the friction coefficient values. 

Unlike the results for the Markovian where the optimal protocol is given by a differential equation, the inclusion of memory leads to an integral equation for the driving in either description. 

It has been provided explicit calculations for the entropy rate of non-Markovian dynamics extending those of Ref. \cite{BroeckEsposito} and reinforcing the findings of Ref. \cite{StrasbergEsposito}. 

It was found the protocol is the only source the system to show a negative entropy rate which clearly violates the second law. It is due to consider in the formulation of the theory that the effects of the degrees of freedom of the bath are described as a unique function of the temperature and as a consequence putting aside important contributions to the total entropy balance.

A reformulation of the FDT is called upon to make the GLE approach thermodynamically consistent.

 %-----------------------------------------------------
%---------------------------------------------------- 
%\section{Final remarks}
%\label{Sec3}
%-----------------------------------------------------
%-----------------------------------------------------

%\section*{Acknowledgments}
%The author thanks the anonymous reviewers for their valuable comments and also Prof. Roy Little of the Universidad de Los Andes for his helpful suggestions.

%The authors thank Prof. Cesare Olinto Colasante for suggestions that greatly improved the manuscript. %We are also grateful for the insightful comments offered by the anonymous peer reviewers.

%---------------------------------------------------------
%---------------------------------------------------------
%\appendix*
%\renewcommand{\theequation}{A.\arabic{equation}}
%\setcounter{equation}{0}
%\section{Title}
%To be written if it is necessary.

%-----------------------------------------------------
%-----------------------------------------------------
\bibliographystyle{apsrev4-1}
\bibliography{references}

\begin{thebibliography}{41}
\bibitem [{Mat()}]{Math}%
  \BibitemOpen
  \href@noop {} {\enquote {\bibinfo {title} {Mathematica{\texttrademark}
  11.2.0.0 computer package was used for the algebraic, numerical and graphics
  manipulations},}\ }\BibitemShut {NoStop}%
\end{thebibliography}
%\nopagebreak
%-----------------------------------------------------
\end{document}